\documentstyle[epsfig,prl,aps,tighten,twocolumn]{revtex}
\newcommand{\be}{\begin{equation}}
\newcommand{\ee}{\end{equation}}
\newcommand{\bea}{\begin{eqnarray}}
\newcommand{\eea}{\end{eqnarray}}
\newcommand{\ba}{\begin{array}}
\newcommand{\ea}{\end{array}}
\newcommand{\nn}{\nonumber}
\newcommand{\PP}{{\cal P}}
\newcommand{\E}{{\cal E}}
\newcommand{\LG}{{\cal L}}
\newcommand{\Th}{\Theta}
\newcommand{\al}{\alpha}
\newcommand{\sech}{\mbox{\rm sech}}
\newcommand{\noi}{\noindent}
\begin{document}
\draft
\title{\bf Spatiotemporally Localized Multidimensional Solitons in 
Self-Induced Transparency Media}

\author{M. Blaauboer,$^{\rm a,b}$ B.A. Malomed,$^{\rm a}$
G. Kurizki, $^{\rm b}$ } 

\address{
$^a$ Department of Interdisciplinary Studies, Faculty of Engineering, 
Tel Aviv University, Tel Aviv 69978, Israel\\
$^b$ Chemical Physics Department, Weizmann Institute of Science, 
Rehovot 76100, Israel
}
\date{\today}
\maketitle

\begin{abstract}
"Light bullets" are multi-dimensional solitons which are localized in both
space and time. We show that such solitons exist in two- and
three-dimensional self-induced-transparency media and that they are 
fully stable. Our approximate analytical calculation, backed 
and verified by direct numerical simulations, yields the multi-dimensional 
generalization of the one-dimensional Sine-Gordon soliton.
\end{abstract}

\pacs{PACS numbers: 42.50 Rh, 42.50 Si, 3.40 Kf
{\tt physics/0002006}}

The concept of multi-dimensional solitons that are localized in 
both space and time, alias "light bullets" (LBs), was pioneered 
by Silberberg\cite{silb90}, and has since then been investigated in 
various nonlinear optical media, with particular emphasis on the question of 
whether these solitons are stable or not. For a second-harmonic generating
medium, the existence of stable two- and three-dimensional (2D and 3D) solitons 
was predicted as early as in 1981\cite{kana81}, followed by studies of 
their propagation and stability against collapse\cite{haya93,malo97,mich98,he98},
and of analogous 3D quantum solitons\cite{kher98}.
In a nonlinear Schr\"odinger model 
both stable and unstable LBs were found\cite{fran98} 
and it was suggested that various models describing
fluid flows yield stable 2D spatio-temporal 
solitons\cite{gott98}. Recently, the first experimental observation 
of a quasi-2D "bullet" in a 3D sample was reported in Ref.~\cite{liu99}.

In this letter we predict a new, hitherto unexplored type of
LBs, obtainable by 2D or 3D {\it self-induced transparency} (SIT).
SIT involves the solitary propagation of an electromagnetic pulse
in a near-resonant medium, irrespective of the carrier-frequency
detuning from resonance\cite{mcca67,lamb71}. The SIT soliton 
in 1D near-resonant media\cite{maim90} is exponentially localized and stable. 
In order to investigate the existence of "light bullets" in SIT, 
i.e. solitons that are localized in both space and time, one has to consider 
a 2D or 3D near-resonant medium. Here we present an approximate analytical solution 
of this problem, which is checked by and in very good agreement with
direct numerical simulations. 

Our starting point are the two-dimensional SIT equations in dimensionless
form\cite{newe92}
\begin{mathletters}
\bea
-i \E_{xx} + \E_{z} - \PP & = & 0 \label{eq:SIT21} \\
\PP_{\tau} - \E W  & = & 0 \label{eq:SIT22} \\
W_{\tau} + \frac{1}{2} (\E^{*} \PP + \PP^{*} \E) & = & 0. \label{eq:SIT23}
\eea
\label{eq:SIT2}
\end{mathletters}

Here $\E$ and $\PP$ denote the slowly-varying amplitudes of the electric field and 
polarization, respectively, $W$ is the inversion, $z$ and $x$ are respectively 
the longitudinal and transverse coordinates (in units of the effective 
absorption length $\al_{\rm eff}$), and $\tau$ the retarded time (in units 
of the input pulse duration $\tau_{p}$). The Fresnel number $F$ ($F > 0$),
which governs the transverse diffraction in 2D and 3D propagation, 
is incorporated in $x$ and the detuning $\Delta \Omega$ of 
the carrier frequency from the central atomic resonance frequency 
is absorbed in $\E$ and $\PP$\cite{trans}. 
We have neglected polarization dephasing and inversion decay, 
considering pulse durations that are much shorter than the corresponding 
relaxation times. Eqs.~(\ref{eq:SIT2}) are then compatible with the local constraint 
$|\PP|^2 + W^2 = 1$, which corresponds
to conservation of the Bloch vector\cite{newe92}.

The first nontrivial question is to find a Lagrangian representation for
these 2D equations, which is necessary for adequate understanding of the
dynamics. To this end, we rewrite the equations in a different form,
introducing  the complex variable $\phi $ defined as follows \cite{lamb73} 
\begin{equation}
\phi \equiv \frac{1+W}{\PP}=\frac{\PP^{\ast }}{1-W}\
\Longleftrightarrow \PP=\frac{2\phi ^{\ast }}{\phi \phi ^{\ast }+1}%
,W=\frac{\phi \phi ^{\ast }-1}{\phi \phi ^{\ast }+1}.  \label{eq:trans1}
\end{equation}
Eqs.~(\ref{eq:SIT22}) and (\ref{eq:SIT23}) can then be expressed as a single
equation, $\phi _{\tau }+(\E/2)\phi ^{2}+(1/2)\E^{\ast }=0$%
. Next, we define a variable $f$ so that $\phi \equiv 2f_{\tau }/(\E f)
$. In terms of $f$, the previous equation becomes $f_{\tau \tau }-
(\E_{\tau }/\E)f_{\tau }+(1/4)|\E|^{2}f=0.$ This equation is
equivalent to 
\begin{mathletters}
\begin{eqnarray}
f_{\tau } &=& \frac{1}{2} \E g  \label{eq:trans3a} \\
g_{\tau } &=& - \frac{1}{2} \E^{\ast}f,  \label{eq:trans3b}
\end{eqnarray}
\label{eq:trans3} 
\end{mathletters}
\noi with $g\equiv f\,\phi $. Applying the same transformations 
to Eq.~(\ref{eq:SIT21}) yields 
\begin{equation}
-i\E_{xx}+\E_{z}-2fg^{\ast }=0.  
\label{eq:trans3c}
\end{equation}
The Lagrangian density corresponding to Eqs.~(\ref{eq:trans3}) and 
(\ref{eq:trans3c}) can now be found in an explicit form,
\bea
\LG(x,\tau) & = & \frac{1}{4} \E_{x} \E_{x}^{*} + \frac{i}{8} 
( \E \E_{z}^{*} - \E_{z} \E^{*}) - \frac{i}{2} \left( f^{*} g \E - 
f g^{*} \E^{*} \right) \nn \\
& &
- \frac{i}{2} \left( f \dot{f}^{*} - \dot{f} f^{*} \right) - 
\frac{i}{2} \left( g \dot{g}^{*} - \dot{g} g^{*} \right).
\label{eq:Lagr}  
\eea

Now we proceed to search for LB solutions. Before resorting to direct
simulations, we obtain an analytical approximation of the solutions.
The starting point for this approximation is the well-known soliton solution
for 1D SIT (the Sine-Gordon soliton)\cite{lamb71,newe92,agra92} 
\begin{mathletters}
\bea
\E(\tau,z) & = & \pm 2\, \al\, \sech \Th 
\label{eq:SGsolE} \\
\PP(\tau,z) & = & \pm 2\, \sech \Th \tanh \Th 
\label{eq:SGsolP} \\
W(\tau,z) & = & \sech^2 \Th - \tanh^2 \Th, 
\label{eq:SGsolW}
\eea
\label{eq:SGsol}
\end{mathletters}
\noi with $\Th(\tau,z) = \al \tau - \frac{z}{\al} + \Th_{0}$, and $\al$, 
$\Th_{0}$  arbitrary real parameters. Equation~(\ref{eq:SGsolE}) 
is also called a $2\pi$-pulse, because 
its area $\int_{\infty}^{\infty} \E(\tau,z) d\tau = \pm 2 \pi$. 

Returning to the 2D SIT equations, we notice by straightforward
substitution into Eqs.~(\ref{eq:trans3}) that a 2D solution with
separated variables, in the form $\E(\tau ,z,x)=\E%
_{1}(\tau ,z)\,\E_{2}(x)$ (and similarly for $f$ and $g$), does 
not exist. To look for less obvious solutions, 
we first split equations~(\ref{eq:SIT2}) into their real and 
imaginary parts, writing $\E \equiv \E_{1} + i \E_{2}$ and 
$\PP \equiv \PP_{1} + i \PP_{2}$:
\begin{mathletters}
\bea
\E_{2xx} + \E_{1z} - \PP_{1} & = & 0 \label{eq:reim11}\\
\E_{1xx} - \E_{2z} + \PP_{2} & = & 0 \label{eq:reim12}\\
\PP_{1\tau} - \E_{1}\, W & = & 0 \label{eq:reim13}\\
\PP_{2\tau} - \E_{2}\, W & = & 0 \label{eq:reim14}\\
W_{\tau} + \E_{1} \PP_{1} + \E_{2} \PP_{2} & = & 0. 
\label{eq:reim15}
\eea
\label{eq:reim1}
\end{mathletters}
In the absence of the $x$-dependence, these equations are 
invariant under the transformation $(\E_{1},\PP_{1})
\leftrightarrow (\E_{2},\PP_{2})$. 
This suggests a 1D solution in which real and imaginary parts
of the field and polarization are equal, $\E_{1} = \E_{2}$ and 
$\PP_{1} = \PP_{2}$, and such that the total field and polarization
reduce to the SG solution~(\ref{eq:SGsol}). Our central 
result is an approximate but quite accurate (see below) extension
of this solution, applicable to the 2D SIT equations. 
In terms of the original physical variables it is given by
\begin{mathletters}
\bea
\E(\tau,z,x) & = & \pm 2 \al \, \sqrt{\sech \Th_{1} \sech \Th_{2}} 
\ \mbox{\rm exp}(- i \Delta \Omega \tau + i \pi / 4) \label{eq:sol21} \\
\PP(\tau,z,x) & = & \pm \sqrt{\sech \Th_{1} \sech \Th_{2}} \{
(\tanh \Th_{1} + \tanh \Th_{2})^2 + \nn \\
& & \frac{1}{4} \al^2 C^4 [ (\tanh \Th_{1}  
- \tanh \Th_{2})^2 - \nn \\
& &  2 (\sech^2 \Th_{1} + \sech^2 \Th_{2}) ]^2 
\}^{1/2}\ \mbox{\rm exp}(- i \Delta \Omega \tau + i \mu) \\
W(\tau,z,x) & = & [1 - \sech \Th_{1} \sech \Th_{2} 
\{ (\tanh \Th_{1} + \tanh \Th_{2})^2 + \nn \\
& & \frac{1}{4} \al^2 C^4 [ (\tanh \Th_{1} - 
\tanh \Th_{2})^2 - \nn \\
& &  2 (\sech^2 \Th_{1} + 
\sech^2 \Th_{2}) ]^2 \} ]^{1/2},
\eea
\label{eq:sol2}
\end{mathletters}
with
\bea
\Th_{1} & = & \al \tau - \frac{z}{\al} + \Th_{0} + C x \nn \\
\Th_{2} & = & \al \tau - \frac{z}{\al} + \Th_{0} - C x, \nn \\
\mu & \equiv & \arctan \left( \PP_{2}/ \PP_{1} \right) \nn.
\eea
Here $\al$, $\Th_{0}$ and $C$ are real constants. 
Equations~(\ref{eq:sol2}) satisfy the two-dimensional
SIT equations~(\ref{eq:reim11}) and (\ref{eq:reim12}) 
and obey the normalization condition $\PP_{1}^2 + \PP_{2}^2 + W^2 = 1$. They 
reduce to the Sine-Gordon solution for $C=0$. The accuracy to which
Eqs.~(\ref{eq:sol2}) satisfy Eqs.~(\ref{eq:reim13})-(\ref{eq:reim15})
is O($\alpha C^2$), which requires that $|\al| C^2\ll 1$. 
This is the single approximation made. Numerical simulations 
discussed later on verify that Eq.~(\ref{eq:sol2}) indeed
approximates the exact solution of Eq.~(\ref{eq:reim1}) to a high 
accuracy. In addition, we have checked that substitution of (\ref{eq:sol2})
into the Lagrangian~(\ref{eq:Lagr}) and varying the resulting expression with respect
to the parameters $\al$ and $C$ yields zero. This "variational approach"
is commonly used to obtain an approximate "ansatz" solution to a set of 
partial differential equations in Lagrangian representation\cite{vari}. 
Equations~(\ref{eq:sol2}) represents a {\it light bullet}, which decays both in space 
and time and is stable for all values of $z$. The latter follows directly 
from (\ref{eq:sol21}) and also from the Vakhitov-Kolokolov stability 
criterion\cite{vaki}.
%
%
\begin{figure}
\centerline{\epsfig{figure=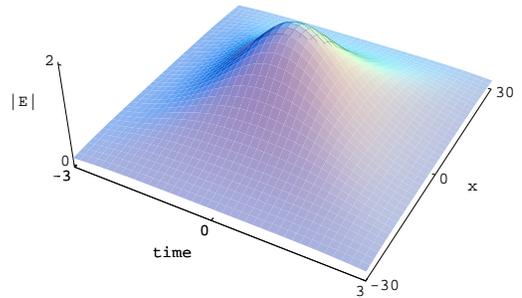,width=0.8\hsize}\vspace{0.5cm}}
\caption[]{The electric field in the 2D "light bullet", $|\E|$, 
as a function of time $\tau$ (in units of the input pulse duration 
$\tau_{p}$) and transverse coordinate $x$ (in  units of the effective 
absorption length $\al_{\rm eff}$) 
after propagating the distance $z=1000$. Parameters used 
correspond to $\al=1$, $C=0.1$ and $\Th_{0}=1000$.
}
\label{fig:field}
\end{figure}
\begin{figure}
\centerline{\epsfig{figure=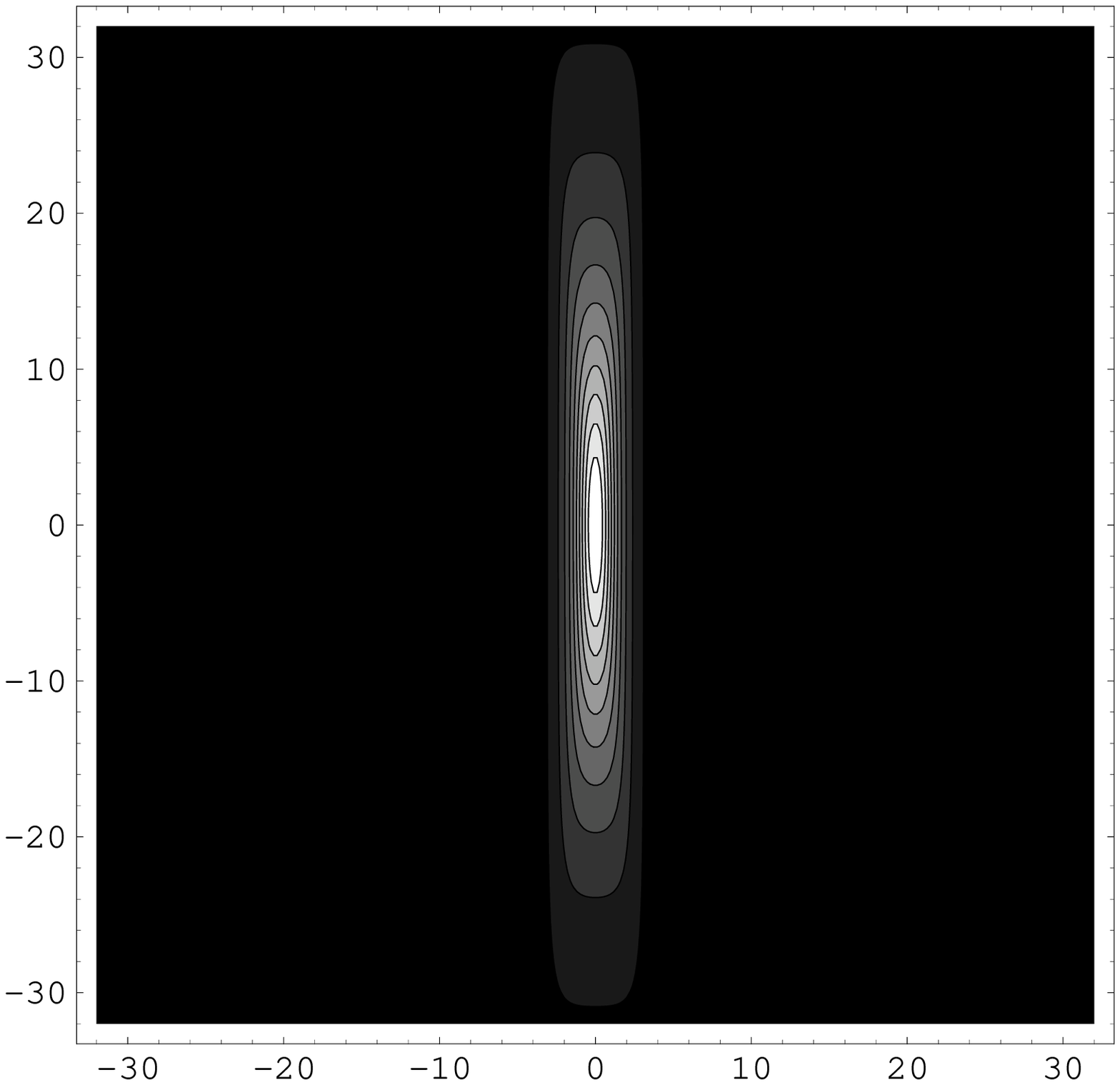,width=0.8\hsize}\vspace{0.5cm}}
\caption[]{Contourplot of Fig.~\ref{fig:field} in the ($\tau$,$x$)-plane.
Regions with lighter shading correspond to higher values of the electric field.
Note the different time scale than that of Fig.~\ref{fig:field}.
}
\label{fig:fieldcont}
\end{figure}
\begin{figure}
\centerline{\epsfig{figure=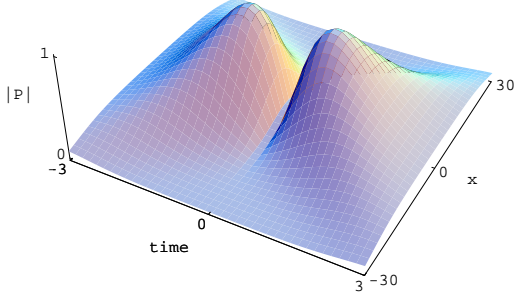,width=0.8\hsize}\vspace{0.5cm}}
\caption[]{The polarization in the 2D "bullet", $|\PP|$, as a function of 
time $\tau$ and transverse coordinate $x$.
Parameters used are the same as in Fig.~\ref{fig:field}.
}
\label{fig:pola}
\end{figure}
Figs.~\ref{fig:field}-\ref{fig:pola} show the electric field and
polarization, generated by direct numerical simulation of the 2D SIT
equations (\ref{eq:SIT2}) at the point $z=1000$, using (\ref{eq:sol2}) as an
initial ansatz for $z=0$. To a very good accuracy (with a deviation $<1\%$),
they still coincide with the initial configuration and analytic
prediction (\ref{eq:sol2}). The electric field has a
typical shape of a 2D LB, localized in time and the transverse coordinate $x$, 
with an amplitude $2\alpha$ and a nearly sech-form cross-section
in a plane in which two of the three coordinates $\tau $, $z$ and $x$ are
constant. The ratio $C/\alpha $ determines how fast the field decays in the
transverse direction. For $|C/\alpha |\ll 1$ (then $|C|<1$, as $|\alpha
|C^{2}\ll 1$), we have a relatively rapid decay in $\tau $ and slow fall-off
in the $x$-direction, as is seen in Fig.~\ref{fig:field}. In the opposite case, 
$|C/\alpha |\gg 1$, the field decays more slowly in time and faster in $x$. The
polarization field has the shape of a double-peaked bullet. Its
cross-section at constant $x$ displays a minimum at $\Theta _{\mbox{\rm min}%
}\approx 0$, where $|\PP(\Theta _{\mbox{\rm min}})|\approx 0$, and maxima at 
$\Theta _{\pm }=\pm \mbox{\rm Arcosh}(\sqrt{2})$, where $|\PP%
(\Theta _{\pm })|\approx 1$. The field and polarization decay in a similar
way, which is a characteristic property of SIT \cite{newe92}. Also the inversion
decays both in time and in $x$, but to a value of $-1$ instead of zero,
corresponding to the atoms in the ground state at infinity.
A numerical calculation of the field area at $x=0$ yields 
$\int_{-\infty }^{\infty }d\tau |\E(\tau ,z,0)| = 6.28 \pm 0.05 
\approx 2\pi $, irrespective of $z$. By analogy with the 
SG soliton, one might thus name this a "$2\pi $ bullet". 

\begin{figure}
\centerline{\epsfig{figure=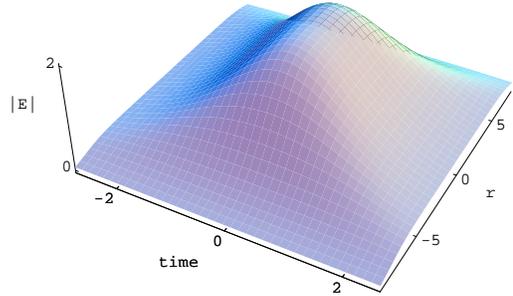,width=0.8\hsize}\vspace{0.5cm}}
\caption[]{The electric field in the 3D "light bullet", $|\E|$, as a function of 
time $\tau$ and transverse radial variable $r$ after propagating
the distance $z=1000$. Parameters used correspond to $\al=1$, 
$C=0.1$ and $\Th_{0}=1000$.
}
\label{fig:E3D}
\end{figure}

We have also numerically obtained {\it axisymmetric} stable LBs in a 3D SIT
medium, see Fig.~\ref{fig:E3D}. The
3D medium is described by Eqs.~(\ref{eq:SIT2}) with the first one replaced
by 
\begin{equation}
-i(\E_{r r }+ r ^{-1}\E_{r })+\E_{z}- 
\PP=0,  \label{eq:SIT3}
\end{equation}
where $r \equiv \sqrt{x^{2}+y^{2}}$ is the transverse radial coordinate.
Searching for an analytic 3D bullet solution in the
transverse plane proves to be difficult. However, in the limit of either
large or small $r$, an approximate analytic solution may be found. For large $r$, 
it again takes the form (\ref{eq:sol2}), but now with $\Theta
_{1}=\alpha \tau -z/\alpha +\Theta _{0}+Cr $ and $\Theta _{2}=\alpha \tau
-z/\alpha +\Theta _{0}-Cr $, where $\alpha $, $\Theta _{0}$, and $C$ are
constants, $|\alpha |C^{2}\ll 1$, and it is implied $r \gg 1/|C|$. It is
in sufficiently good agreement (deviations $<5\%$) with results of 
simulation of the 3D equations, using this solution as an initial ansatz. 
Comparison of Figs.~\ref{fig:field} and
~\ref{fig:E3D} shows that the 2D and 3D bullets have similar shapes, but the
3D one decays faster in the radial direction for small $r$ than the 2D
bullet in its transverse direction.

For constant $\tau$, the 2D and 3D bullets 
are localized in both the propagation direction $z$ and the transverse 
direction(s). One may also ask whether there exist SIT solitons which 
are traveling (plane) waves in $z$ and localized in $x$ (and $y$).
Using a symmetry argument, it is straightforward to prove that 
they do {\it not} exist. Starting from the SIT equations (\ref{eq:SIT2})
(in 2D, the 3D case can be considered analogously) we adopt a plane-wave ansatz
for $\E$ and $\PP$, changing variables as follows: 
$x\rightarrow \sqrt{k}x$ (assuming $k>0
$), $\E(\tau ,z,x)\rightarrow \E(\tau ,x)\,\exp
(-ikz)$, $\PP(\tau ,z,x)\rightarrow k^{-1}\,%
\PP(\tau ,x)\,\exp (-ikz)$, and $W(\tau
,z,x)\rightarrow k^{-1}\,W(\tau ,x)$. The equations for the real and
imaginary parts of the field then become 
\begin{mathletters}
\bea
\E_{2xx} - \E_{2} - \PP_{1} & = & 0 \label{eq:reim31}\\
\E_{1xx} - \E_{1} + \PP_{2} & = & 0, \label{eq:reim32}
\eea
\label{eq:reim3}
\end{mathletters}
\noi with the equations for $\PP_{\tau}$ and $W_{\tau}$ given by 
(\ref{eq:reim13})-(\ref{eq:reim15}). Using the transformation 
$(\E_{1},\PP_{1}) \leftrightarrow (\E_{2},\PP_{2})$,
which leaves the last three equations invariant but changes the first two,
one immediately finds that (\ref{eq:reim3}) only admits the trivial
solution $\E_{1}=\E_{2}=\PP_{1}=\PP_{2}=0$, $W=-1$.

The observation of "light bullets" in a SIT process 
requires high input power of the incident pulse 
and high density of the two-level atoms in the medium, in order to 
achieve pulse durations short compared to decoherence and loss times. 
These requirements are met e.g. for alkali gas media, 
with typical atomic densities of $\sim 10^{11}$ atoms/cm$^3$ and relaxation times
$\sim 50 $ ns\cite{slus74}, and for optical pulses generated by 
a laser with pulse duration $\tau_{p} < 0.1$ ns.
In order to include transverse diffraction, the incident pulse should be 
of uniform transverse intensity and satisfy $\al_{\rm eff}d^2/\lambda < 1$
\cite{slus74}, where $\lambda$ and $d$ are its carrier wavelength and diameter 
respectively\cite{slus74}. The parameter $\alpha$ in the solution
(\ref{eq:sol2}), which determines the amplitude of the bullet
and its decay in time, corresponds to $\alpha \sim \kappa_{z} 
\tau_{p} v_{p}$\cite{maim90}, with $\kappa_{z}$ the wavevector component along
the propagation direction $z$ and $v_{p}$ the velocity of the 
pulse in the medium, and can thus be controlled by the incident pulse duration and
velocity. The parameter $C \sim \kappa_{x} L_{x}$, where $\kappa_{x}$ is the transverse
component of the wavevector and $L_{x}$ is the spatial transverse width
of the pulse, is also controlled by the characteristics
of the incident pulse and should satisfy the condition $\kappa_{z} \kappa_{x}^2
L_{z} L_{x}^2 \ll 1$. For a homogeneous (atomic beam) absorber, the effective absorption length 
$\alpha_{\rm eff} \sim 10^4$ m$^{-1}$ and the Fresnel number $F$ can range from 
1 to 100\cite{slus74}. The bullets then decay on a time scale 
of $t\sim 1-10\, \tau_{p} \sim 10$ ns and transverse length of
$x \sim 0.1-1$ mm, which is well 
within experimental reach.

In conclusion, we predict the existence of fully stable "light bullets"
in 2D and 3D self-induced transparency media. The prediction
is based on an approximate analytical solution of the 
multi-dimensional SIT equations and verified by direct numerical 
simulation of these PDE's. Our results suggest an experiment aimed at detection
of this "bullet" in an SIT-medium and opens the road for 
analogous searches for "light bullets" in other nonlinear optical processes,
such as, e.g., stimulated Raman scattering, which is analogous to SIT.

M.B. acknowledges support from the Israeli 
Council for Higher Education. Support from ISF, Minerva and EU (TMR)
is acknowledged by G.K.

\end{document}